\documentclass[preprint2]{aastex}
\newcommand{\Hb}{H$\beta$}

\newcommand{\kms}{\,km\,s$^{-1}$}
\newcommand{\HI}{\ion{H}{1}}
\newcommand{\HII}{\ion{H}{2}}
\newcommand{\OIII}{[\ion{O}{3}]}

\newcommand{\dV}{$\Delta V_\mathrm{max}$}
\newcommand{\AM}{AM\,1353$-$272}
\shorttitle{Velocity gradients in \AM}
\shortauthors{Weilbacher et al.}
\begin{document}
\title{
  Large Velocity Gradients in the Tidal Tails of the\\
  Interacting Galaxy \AM\ (``The Dentist's Chair'')
  \altaffilmark{1}
}
\author{
  Peter M.~Weilbacher\altaffilmark{2},
  Uta Fritze-v.~Alvensleben\altaffilmark{2},
  Pierre-Alain Duc\altaffilmark{3},
  Klaus J.~Fricke\altaffilmark{2}
}
\altaffiltext{1}{Based on observations collected at the European Southern Observatory, La Silla, Chile (ESO No 67.B-0049).}
\altaffiltext{2}{Universit\"ats-Sternwarte, Geismarlandstra\ss{}e 11, 37083 G\"ottingen, Germany, \{weilbach,ufritze\}@uni-sw.gwdg.de}
\altaffiltext{3}{CNRS URA 2052 and CEA, DSM, DAPNIA, Service d'Astrophysique, Centre d'Etudes de Saclay, 91191 Gif-sur-Yvette Cedex, France, paduc@cea.fr}
\begin{abstract}
  We present VLT observations of the interacting system \AM. Using the
  FORS2 instrument, we studied the kinematics of the ionized gas along
  its prominent tidal tails and discovered strikingly large velocity
  gradients associated with seven luminous tidal knots.  These
  kinematical structures cannot be caused by streaming motion and most
  likely do not result from projection effects.
  More probably, instabilities in the tidal tails have lead to the
  formation of kinematically decoupled objects which could be the
  progenitors of self-gravitating Tidal Dwarf Galaxies.
\end{abstract}
%
%
\section{Introduction}
Interactions among disk galaxies are observed to produce tidal tails
of sometimes impressive lengths containing stars and HI in varying
proportions.
While tidal tails of interacting galaxies have been analyzed in
dynamical simulations of interactions to assess their use to probe the
form of the dark matter potential \citep[e.g.][]{SW99}, the detailed
kinematics of structures within the tidal tails have not received much
attention.
Both in deep spectroscopic observations \citep{DBS+00} and high
resolution dynamical simulations \citep{BH92}, condensations are
observed to form in tidal tails, the nature and fate of which are not
yet fully understood.

At a distance of 159~Mpc (computed with $H_0 = 75$\kms\,Mpc$^{-1}$)
the interacting system \objectname[AM 1353-272]{\AM}, nicknamed the
``The Dentist's Chair'' for its peculiar morphology, consists of three
apparent components (see Fig.~\ref{fig:chart}): `A', a disturbed
galaxy with two $\sim$40\,kpc long tidal tails (also cataloged as
ESO\,510-G\,020G and IRAS\,F13533-2721), a disturbed edge-on disk
galaxy `B', and an elliptical `C' \citep[see][]{WDF+00}.
While `A' and `B' are a physical pair with a central velocity
difference of only $\sim 150$\kms, \objectname[AM 1353-272C]{`C'} has
a heliocentric velocity of 14750\kms, and is therefore located 38\,Mpc
behind the interacting pair.  The tidal tails of \AM\,A host several
blue knots with luminosities of dwarf galaxies. From their location
and colors, they were identified by \citet{WDF+00} as Tidal Dwarf
Galaxy (TDG) candidates.

In this letter, we present the first evidence from deep optical
spectroscopy with the VLT that some dense structures in the tidal
tails are kinematically decoupled from the overall motion of the
tails.  Forthcoming instruments on large telescopes --- e.g.~high
resolution integral field spectrographs --- will be very well suited
to investigate further the specific dynamics in the tidal tails of
\AM\,A.
%
%
\section{Observations and data reduction}\label{sec:obsData}
\begin{figure}
  \plotone{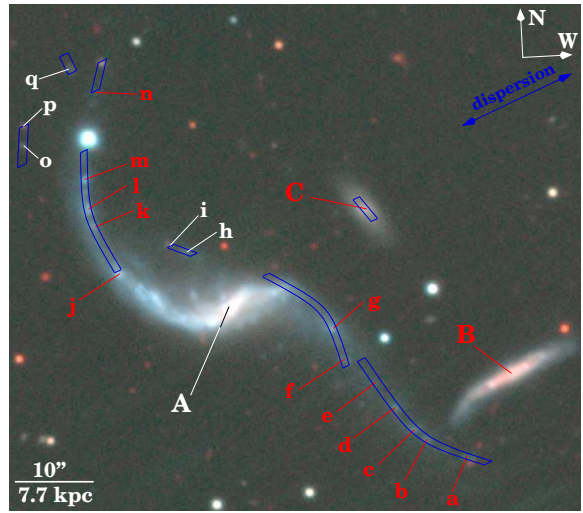}
  \caption{Finding chart of \AM.
    Original $B$-band image from NTT-SUSI \citep{WDF+00} in logarithmic scale.
    (Electronic edition: Composite $B$+$V$+NIR ``true'' color image from NTT-SUSI and SOFI \citep[][Weilbacher et al.~in prep.]{WDF+00}.)
    The field of view is $\sim$2\arcmin$\times$2\arcmin.
    Black labels (red labels in electronic edition) mark objects with
    measured redshift.
    \label{fig:chart}
  }
\end{figure}

We have obtained multi-object spectroscopic data of a field centered
on the system \AM\ with the FORS2 instrument at the 4th VLT telescope
``Yepun''. Service mode observations were carried out in the night
13/14th of August 2001 under photometric conditions.  Several
spectroscopic standards were observed. The total exposure time of
3210\,s was split into three individual exposures of 1070\,s to ease
cosmic ray cleaning.  The seeing of 1\farcs0 is well sampled by an
instrumental scale of 0\farcs2\,px$^{-1}$.  The grism 600B+22 was used,
giving a wavelength coverage of 3450\dots5900\,\AA\ and a spectral
resolution of 5.7\,\AA\ (1.2\,\AA\ per pixel) for a central slit of
1\arcsec\ width.  We created a mask for the mask exchange unit (MXU)
of the FORS2 instrument with curved and tilted slits to cover the
tails of \AM\,A and several surrounding galaxies in one setup (see
Fig.~\ref{fig:chart}).

\begin{figure}
  \plotone{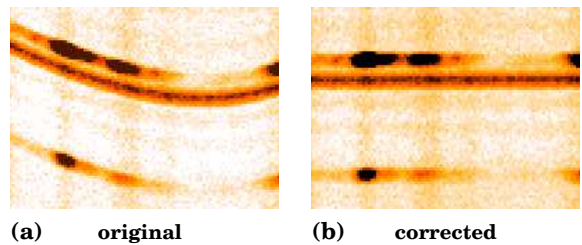}
  \caption{Slit curvature correction.
    As an example, we show part of a 2D spectrum with the emission lines
    \OIII5007,4959 emitted by the knots `j' to `m'. Between these object
    lines, a skyline is visible which allows to judge the quality of the
    correction.
    \label{fig:corCurve}
  }
\end{figure}

The data reduction followed standard recipes, for which we used our
own IRAF task \texttt{mosx} \citep{WDFA02}.  Special care was taken to
correct the curvature of the 2D spectra of the curved slits: a proper
wavelength calibration was carried out at each position along the slits,
using a reference HeHgCd frame observed through the same mask as the
science frames.  The typical accuracy for each column is 0.04\,\AA\ RMS. A
4th order polynomial fit was then performed for each slit to determine
the final wavelength solution and compute the parameters of the curvature
correction. All these tasks were performed using the standard tools in
the IRAF package \texttt{longslit}.
The result of this procedure is presented in Fig.~\ref{fig:corCurve}:
the original, uncorrected 2D spectrum is compared with the spectrum
after the correction.

After subtracting the sky background, we used the brightest emission lines
from the slits along the tidal tails to derive the velocity profiles with
an IRAF script based on the \texttt{fitprofs} procedure. As a check,
the same velocity fit was carried out with the nearest lines in the
wavelength calibration spectrum.
The 1$\sigma$ differences from zero velocity in this control fit are
below 1\kms.  The {\it relative systematic} errors in the final velocity
profiles should be of the order of 1\kms\ while the errors for single
points are typically 15\kms.
%
%
\section{Results}
\begin{figure*}
  \epsscale{1.8}
  \plotone{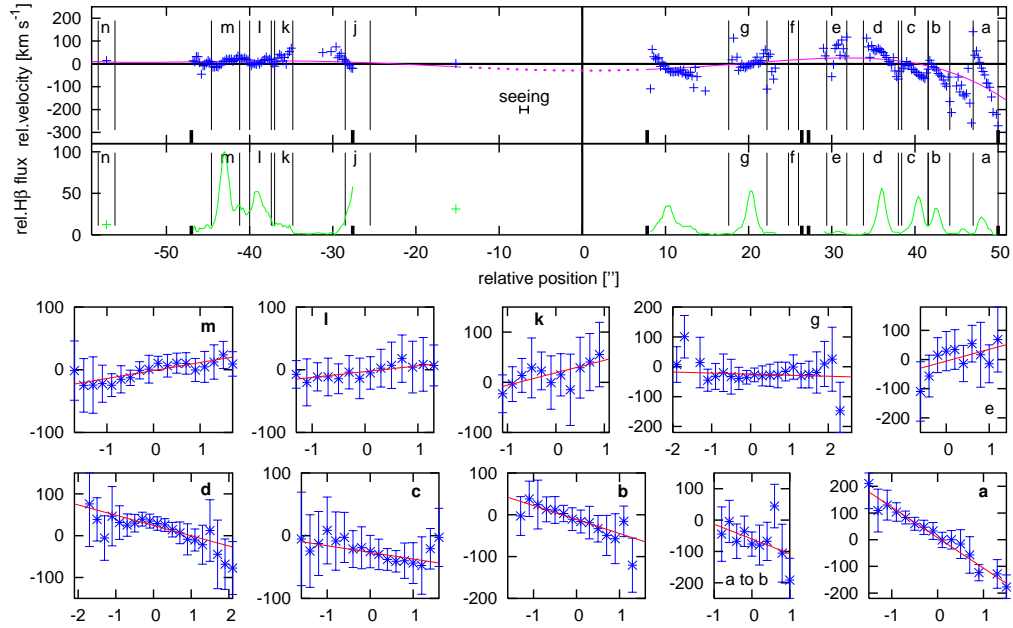}
  \caption{Velocities along \AM\,A. 
    {\bf Top}: velocity field relative to a zeropoint of 11935\kms\ 
    along the ridge of the galaxy from the northern (left) to the
    southwestern (right) tip of the tidal tail. The seeing of 1\farcs0 is
    indicated. The lower part shows the \Hb\ flux in relative units on
    the same spatial scale. The size of each knot is marked with
    vertical lines. Bold marks on the position axis indicate the ends
    of the curved slits. A 5th order fit to the overall velocity profile
    is shown. 
    {\bf Bottom}: the extracted residual velocity field of each knot
    after correction for tidal motion, plotted as relative position
    [\arcsec] vs.~relative velocity [km\,s$^{-1}$]. Lines show a
    linear fit.
    \label{fig:veloplot}
  }
\end{figure*}
Fig.~\ref{fig:veloplot} presents the velocity distribution of the
ionized gas along the two tails of \AM.  The velocities indicated on
the top of Fig.~\ref{fig:veloplot} are relative to the interpolated
central velocity $V_A = 11935$\kms\ of \AM\,A.  The relative \Hb\ flux
and hence distribution of the \HII\ regions along the tails, is shown
below: it may be used to infer the S/N ratio of the emission lines
used to derive the velocities.  Vertical lines mark the individual
optical knots studied by \citet{WDF+00}.  Note that nearly all of them
are sites of active star formation, as predicted by a comparison of
photometric starburst models with optical broad band colors.

At first glance, it seems as if there are velocity gradients within
the knots and sudden jumps in velocity between them. To further
analyze the step-like shape of the velocity curve, we subtract the
overall streaming motion of the tails, by fitting a 5th order cubic
spline (using IRAF's \texttt{curfit} task) to all velocity points.  We
then extract individual regions from the residual velocity field (see
lower part of Fig.~\ref{fig:veloplot}).

\begin{table*}
    \caption{Velocity gradients within the knots.\label{tab:res}}
    \begin{tabular}{ l | rrc  }
      ID\tablenotemark{a}
                   & \multicolumn{1}{c}{\dV\tablenotemark{b}} 
                            & error & extent\tablenotemark{c}          \\
                   & [km\,s$^{-1}$] & & [\arcsec], [kpc]      \\
      \tableline\tableline
      {\bf a}      & $+343$ &   5\% & 3\farcs0, 2.3 \\
      a-to-b\tablenotemark{d}
                   &  $+94$ &  72\% & 2\farcs0, 1.5 \\
      {\bf b}      &  $+87$ &  21\% & 2\farcs6, 2.0 \\
      {\bf c}      &  $+34$ &  40\% & 3\farcs2, 2.5 \\
      {\bf d}      &  $+93$ &  10\% & 4\farcs2, 2.9 \\
      e            &  $-73$ &  65\% & 2\farcs4, 1.4 \\
      f            & \multicolumn{3}{c}{\dots\dots\ \ low S/N\ \ \dots\dots} \\
      g            &  $+15$ & 170\% & 4\farcs6, 3.2 \\
      \tableline
      {\bf k}      &  $-50$ &  31\% & 2\farcs2, 1.7 \\
      {\bf l}      &  $-24$ &  24\% & 2\farcs6, 2.0 \\
      {\bf m}      &  $-43$ &  19\% & 3\farcs4, 2.6 \\
      \tableline
    \end{tabular}
    \tablenotetext{a}{Knots with significant amplitude of velocity
      gradient are marked in boldface.}
    \tablenotetext{b}{Gradients with positive sign have the highest
      velocity towards the north of the system.}
    \tablenotetext{c}{The seeing of 1\arcsec\ at the distance of
      \AM\,A corresponds to $\sim$800\,pc.}
    \tablenotetext{d}{We designate the region $\pm$1\arcsec\ around
      +45\farcs8 distance from the nucleus as a-to-b.}
\end{table*}

In order to quantify the remaining apparent velocity gradients, we
tried to fit a linear relation to all velocity gradients within the
spatial extent of the knots.  The results are given in
Table~\ref{tab:res}, together with the error given by the fit
procedure, the maximum velocity difference \dV, and the angular extent
of the fitted gradient. The fits seem to be a good first order
approximation to the real velocity gradients in most cases, and are
shown in the bottom part of Fig.~\ref{fig:veloplot}.  Seven knots show
velocity amplitudes that are significant, with errors below 50\%.
They reach 340\kms\ for object `a'.  For several knots the gradient is
also clearly resolved, i.e.~it is observed to extend more than twice
the seeing of 1\arcsec.
%
%
\section{Discussion and Conclusions}
Given the observed significant velocity gradients within seven knots of the
tails of \AM\,A, we discuss several possibilities of the origin of
this specific velocity distribution.

Instrumental errors can be excluded as a cause of these gradients. The
velocity profiles observed over very small angular sizes (a few
arcseconds) cannot be explained with flexures or distortions of the
instrument.  The wavelength calibration has been carried out and checked
with the procedure described in Sect.~\ref{sec:obsData}. As a result, any
relative instrumental effects within a given slit should be below 1\kms.

The apparent gradients could result from projection effects.  As the
angular size probed is only about 2 to 4 times the seeing, several
smaller, physically unrelated knots might appear blended into one larger
clump, mimicking the gradient we observe. It seems improbable, however,
that this should be the case for many knots.
Besides the two well determined tidal tails, there is no indication of
complex tidal structures in this system.  It is therefore quite unlikely
that tidal debris outside the well-defined tails is observed in projection
close to so many tidal knots.\\
Projection effects within a single tail are possible. First, the
tails may have some depth along the line of sight, as can be seen in
dynamical models of interactions starting with the idea of \citet{TT72}
to view tails as two-dimensional ribbons.  Second, the very tip of the
southern tail could indeed be bent.  Its three-dimensional shape might
partly account for the exceptionally large velocity gradient observed
towards condensation `a'.  An example of a projection by such a bend
near the end of a tail is discussed by \citet{HvdH+01} for the tail of
the Antennae galaxies (NGC\,4038/39).  There, the total \HI\ column
density and the velocity dispersion along the line of sight mimic a
massive condensation of matter. In their high-resolution \HI\ data,
they find two gas concentrations at different velocities, which have
only about a tenth of the originally estimated mass.
The amplitude \dV\,$=340$\kms\ within knot `a' (compared to
\dV\,$\sim50\dots100$\kms\ in the Antennae region) makes it seem unlikely
to be caused by projection alone.
The overall velocity field towards the end of the southern tail is
decreasing. If the tail was bent backwards to the center, one would expect
this trend to continue. Knots projected from this bent part of the tail
should then have even lower velocity than `a'. Such velocities are not
observed. Projection effects by a bent tail for other knots apart from
`a' are implausible.

Velocity gradients could, in principle, be caused by gaseous outflows
where one side of an expanding shell is blocked by dust.  This would
require significant amounts of extinction within the knots, while we find
only very moderate absorption ($A_B \lesssim 1.0$\,mag including galactic
$A_B=0.26$\,mag, estimated from Balmer line ratios).  Additionally,
an outflow would also imply line-broadening. There is no indication
of broad lines here, as the FWHM of the \Hb\ line is the same as the
spectral resolution.  It is very unlikely that the gradients are caused
by outflows.

Rotation of bound objects could also create this kind of velocity
gradients. From the irregular optical appearance and the location of
the knots within tidal tails, it is unlikely that the stars in the
potential of a knot have stable orbits. True Keplerian motion
will not apply here, and it would then be premature to interpret the
observed kinematics in terms of ``rotation''.  Mass estimates from the
Virial theorem would strongly overestimate the real mass contained within
the knots\footnote{If caused by Keplerian rotation alone, the
  rotational velocity of knot `a' would be of the same order as those of
  giant spiral galaxies, which is unrealistic for an object with $M_B =
  -13.7$\,mag.}.

Instead, we may be witnessing the formation of bound objects in the
tidal tails which are not yet virialized. The velocity gradients also
seem to be specifically oriented in the sense that the higher velocity
end of each knot points towards the center of \AM\,A. A possible cause
could be the rotational direction of progenitor disk before
the interaction. The presence of the companion galaxy `B' near the
southwestern tidal tail could also amplify motions in the tails, to
create a ``spin'' in these knots in one direction. This would
also explain the higher velocity amplitudes within the knots in the
southwestern tail close to the companion.

The idea of Tidal Dwarf Galaxies (TDGs) --- dwarf galaxies that are
born as condensations in tidal tails --- has been discussed in recent
years by several authors \citep{BH92,HM95,DBW+97}, both theoretically
and observationally. \citet{DBS+00} proposed a first definition of the
term TDG as a self-gravitating object of dwarf galaxy mass formed from
tidal material \citep[see also][]{WD01}.  
While we are hampered by the spatial resolution and cannot give the
ultimate proof of the velocity gradients as motion of tidal material
beginning to get bound into knots, it seems that we are witnessing the
formation of TDGs in the tails of \AM\,A, for the first time for multiple
objects along both tails of the same interacting galaxy.

Forthcoming instruments for high-resolution spectroscopy on 8m-class
telescopes, e.g.~integral field instruments, will enable us to confirm
the velocity gradients and further investigate their origin.  Coupled
with the high spatial resolution of adaptive optics these instruments 
will be very useful tools to analyze the unique properties of the knots seen
in these tidal tails.  To fully understand the geometry and
interaction parameters causing this first series of possibly genuine
Tidal Dwarf Galaxies and to complement the observations, detailed
dynamical N-body+SPH modeling of the collision will be necessary.

\acknowledgments
We thank I.~Appenzeller for help with the proposal and data retrieval and
our anonymous referee for a friendly and helpful report. 
PMW is partly supported by DFG grant FR 916/6-2.
\bibliography{/home/leo/weilbach/Texte/PmW}
\end{document}